\documentstyle[aps,eqsecnum,amsfonts,psfig,preprint]{revtex}
\begin{document}
\title{ \hfill OKHEP-95-04\\ \hfill Imperial/TP/94--95/27\\
Non-Abelian Finite-Element Gauge Theory}
\author{Kimball A. Milton\thanks{E-mail: kmilton@uoknor.edu or
k.milton@ic.ac.uk (until 1 July 1995)}}
\address{\it Department of Physics and Astronomy,
The University of Oklahoma, Norman OK 73019, USA\thanks{
Permanent address}
\\
Theoretical Physics Group, Blackett Laboratory,
 Imperial College, Prince Consort Road, London
SW7 2BZ, UK}

\date{\today}
\maketitle

\begin{abstract}
We complete the formulation of the equations of motion of a
non-Abelian gauge field coupled to fermions on a finite-element
lattice in four space-time dimensions.  This is accomplished by
a straightforward iterative approach, in which successive
interaction terms are added to the Dirac and Yang-Mills equations
of motion, and to the field strength, in order to preserve
lattice gauge invariance exactly, yielding a series in powers of
$ghA$.  Here $g$ is the coupling constant, $h$ is the lattice
spacing, and $A$ is the gauge potential.  Gauge transformations
of the potentials are determined simultaneously.  The interaction
terms in the equations of motion are nonlocal, and can be expressed
either by an iterative formula  or by a difference equation.
On the other hand, the field strength is locally constructed
from the potentials in terms of  a path-ordered product of
exponentials.

\end{abstract}
\section{Introduction}
An intriguing approach to quantum field theories, called the finite
element method, has been under development for the past decade.
(For a recent review see \cite{review}.)  This technique of putting
the operator equations of motion on a lattice has a number of
virtues:
\begin{itemize}
\item It is exactly unitary, or canonical, in the sense that the
equal-time commutation relations are preserved at each lattice site.
\item It is the most accurate possible method of discretizing
the continuum equations of motion:  Typically, after $N$ steps
through
the lattice, the relative error is only $O(1/N^2)$.
\item Because it is an analytic technique, it is not subject to
the statistical errors of Monte Carlo methods.
\item It appears to resolve the fermion-doubling problem, while
breaking chiral invariance with the expected axial-vector anomaly
\cite{mmsb2,twodqed}.
\item Although a lattice Lagrangian does not exist in Minkowski
space, it is possible to define a Hamiltonian in the sense of a
time-evolution operator, which can be used to extract spectral and
other information about the theory \cite{coc,hamm}.
\end{itemize}

Very early in this development it was discovered how to formulate
an Abelian gauge theory in the finite-element scheme\footnote{A
rather different approach to the $U(1)$ gauge theory on a
finite element lattice is given in \cite{mat}.} \cite{qed}.
This was a nontrivial accomplishment, because, {\it a priori},
it is not obvious that the variational gauge-covariance equations are
integrable.  It has seemed to be much more difficult to formulate a
non-Abelian theory.  The essential development took place in the
late 1980's \cite{nagt}.  There it was shown, in two space-time
dimensions, how to write down the interacting Dirac equation
and the Yang-Mills equations, where the interactions were determined
simultaneously with the gauge transformations.  The integrability
of this process was again highly nontrivial.  The only thing not
completely determined at that time was the form of the construction
of the field strength in terms of potentials.  (This was given
only out to fourth order in potentials.)

The generalization to  four dimensions seemed straightforward.
The form for the equations of motion was immediately generalizable,
but only the first two terms in the field strength were given
in 1990 \cite{sing}.  Attention shifted to lattice QED \cite{mmsb}.

I returned to this problem in January of this year.  I computed
the first three terms in the field strength in four dimensions, and
then recognized the pattern in two and four dimensions.  The
construction is remarkably simple, and should have been recognized
very early on:  The field strength is simply given in terms of
a path-ordered product of exponentials of the natural finite-element
gauge-potential variable
(link operators), with further link operators providing the necessary
gauge transformations in the transverse directions.
 From the result, verification of
gauge covariance is immediate.

The organization of this paper is very simple. In the next section
we write down the free lattice equations of motion and explain
how they are gauged.  The full form of the Dirac equation, Yang-Mills
equations, and the gauge transformation will be given.  In Section
III the results of the iterative calculation for the field
strength through  order $g^2 h A^3$ are given,
where $g$ is the gauge coupling constant, $h$ is the lattice
spacing, and $A$ is the vector potential.
  In Section IV we infer
the exact form of the field strength, which is immediately
verified to be gauge covariant.  The continuing direction of this
research program is sketched in the concluding Section V\null.
The essential element of the proof of gauge covariance is given
in the Appendix.

\section{Finite-Element Equations of Motion and Gauge Invariance}
The two-dimensional work has been described in detail
\cite{review,nagt}, so we will begin immediately in four-dimensional
Minkowski space.  Let us start from the free Dirac equation in the
continuum,
\begin{equation}
(i\gamma^\mu\partial_\mu+\mu)\psi=0.
\end{equation}
The linear finite-element prescription consists in replacing
derivatives by forward differences, while in directions in which
there is no differentiation, a forward average is employed.
If we use the following notation for forward averaging,
\begin{equation}
x_{\overline{m}}={1\over2}(x_{m+1}+x_m),
\end{equation}
the free finite-element lattice Dirac equation is
\begin{equation}
{i\gamma^0\over h}(\psi_{\overline{\bf m},n+1}-
\psi_{\overline{\bf m},n})+{i\gamma^j\over\Delta}
(\psi_{m_j+1,\overline{\bf m}_\perp,\overline{n}}-
\psi_{m_j,\overline{\bf m}_\perp,\overline{n}})
+\mu\psi_{\overline{\bf m},\overline{n}}=0,
\label{dirac}
\end{equation}
where $h$ is the temporal lattice spacing, $\Delta$ is the
spatial lattice spacing, $\bf m$ represents a spatial lattice
coordinate, $n$ a temporal coordinate, and the $\perp$ indicates
directions other than the one singled out.

Gauge transformations are introduced just as in the continuum.
The Dirac equation (\ref{dirac}) is invariant under the
infinitesimal global phase transformation
\begin{equation}
\delta\psi_{{\bf m},n}=ig\delta\omega\psi_{{\bf m},n}.
\end{equation}
However, if $\delta\omega$ depends on position, invariance is
spoiled.  Let us define a local transformation so that the mass
term in (\ref{dirac}) transforms covariantly:
\begin{equation}
\delta\psi_{\overline{\bf m},\overline{n}}=ig\delta
\omega_{{\bf m},n}\psi_{\overline{\bf m},\overline{n}}.
\label{local}
\end{equation}
The forward differences in (\ref{dirac}) do not transform
covariantly; to achieve covariance we must add interaction
terms.  The interaction is with a vector potential transforming
by the lattice analog of
\begin{equation}
A_\mu\to A_\mu+\delta^{(0)}A_\mu,\quad \delta^{(0)}A_\mu
=\partial_\mu\omega,
\end{equation}
that is,
\begin{equation}
\delta^{(0)}(A_0)_{\text{m}}={1\over h}(\delta
\Lambda_{\overline{\bf m},n+1}
-\delta\Lambda_{\overline{\bf m},n}),\quad
\delta^{(0)}(A_j)_{\text{m}}={1\over\Delta}(\delta\Lambda_{m_j+1,
\overline{\bf m}_\perp,\overline{n}}-\delta\Lambda_{m_j,
\overline{\bf m}_\perp,\overline{n}}),
\label{gradtrans}
\end{equation}
where $\text{m}$ represents the four-vector coordinate
$({\bf m},n)$ and the superscript $(0)$ is a reminder  that
this is the first term in a series of variations.
  We will take the connection between $\delta\omega$
and $\delta\Lambda$ to be the finite-element one:
\begin{equation}
\delta\omega_{\text{m}}=\delta\Lambda_{\overline{\text{m}}}.
\end{equation}
Then, using the boundary conditions (where $M$ is the number of
lattice sites in any spatial direction)
\begin{equation}
\psi_{m_j+M,{\bf m}_\perp,n}
=(-1)^{M+1}\psi_{m_j,{\bf m}_\perp,n},\quad
\delta\psi_{m_j+M,{\bf m}_\perp,n}=(-1)^{M+1}
\delta\psi_{m_j,{\bf m}_\perp,n},
\label{periodic}
\end{equation}
and the initial condition
\begin{equation}
\delta\psi_{\overline{\bf m},0}
={ig\over2}(\delta\omega_{{\bf m},0}
+\delta\omega_{{\bf m},-1})\psi_{\overline{\bf m},0},
\label{initial}
\end{equation}
 it is easy to see that the noncovariance of
\begin{equation}
{i\gamma^0\over h}(\psi_{\overline{\bf m},n+1}
-\psi_{\overline{\bf m},n})
+{i\gamma^j\over\Delta}(\psi_{m_j+1,\overline{\bf m}_\perp,
\overline{n}}
-\psi_{m_j,\overline{\bf m}_\perp,\overline{n}})
\end{equation}
will be cancelled if the following interaction terms are
added to the left side of (\ref{dirac}):
\begin{equation}
I^{(1)}_{\text{m}}={i\gamma^j\over\Delta}ig\Delta
\sum_{m_j'=m_j+1}^{m_j+M}(-1)^{m_j+m_j'}
(A_j)_{\overline{m'_j-1},{\bf m}_\perp,n}
\psi_{m'_j,\overline{\bf m}_\perp,
\overline{n}}
\label{firstspace}
\end{equation}
and
\begin{equation}
K^{(1)}_{\text{m}}=-2{i\gamma^0\over h}igh
\sum_{n'=0}^n\!\!{}'(-1)^{n+n'}
(A_0)_{{\bf m},\overline{n'-1}}
\psi_{\overline{\bf m},n'},
\label{firsttime}
\end{equation}
where the prime on the summation sign indicates that the $n'=0$ term
is counted with half weight.  The variations to which the interaction
terms are subjected are, from (\ref{gradtrans}),
\begin{mathletters}
\label{zerovar}
\begin{eqnarray}
\delta^{(0)}(A_j)_{\overline{m_j-1},{\bf m}_\perp,n}
&=&{1\over \Delta}(\delta\omega_{m_j,{\bf m}_\perp,n}-
\delta\omega_{m_j-1,{\bf m}_\perp,n})\\
\delta^{(0)}(A_0)_{{\bf m},\overline{n-1}}
&=&{1\over h}(\delta\omega_{{\bf m},n}-
\delta\omega_{{\bf m},n-1}).
\end{eqnarray}
\end{mathletters}
It will be noted that the form of the interaction term with the
scalar potential involves only the fields at the current and earlier
times.  This feature persists, and enables one to solve the equations
by time-stepping through the lattice.
It is easy to verify \cite{qed,nagt} that the Dirac equation with
(\ref{firstspace}) and (\ref{firsttime}) included reduces to the
correct continuum limit. [In this verification, the form of the
initial
condition (\ref{initial}) is crucial.]

However, we are not finished, because we have not varied the
interaction
terms (\ref{firstspace}) and (\ref{firsttime}) with respect to
$\psi$,
(\ref{local}), nor achieved covariance of the interaction term.
This can be accomplished by adding  new interaction terms $I^{(2)}$
and
$K^{(2)}$, and additional gauge transformations, denoted by
$\delta^{(1)}$.
A straightforward calculation reveals that
\begin{equation}
\delta_\psi
I^{(1)}_{\text{m}}=ig\delta\omega_{\text{m}}I^{(1)}_{\text{m}}
-\delta^{(1)}I^{(1)}_{\text{m}}-\delta^{(0)}I^{(2)}_{\text{m}}.
\end{equation}
Here the first term on the right
is the required covariance term, the second
term involves a new variation of ${\bf A}$,
\begin{equation}
\delta^{(1)}(A_j)_{\overline{m'_j-1},{\bf m}_\perp,n}
=ig{1\over2}[\delta\omega_{m_j,{\bf m}_\perp,n}+
\delta\omega_{m_j-1,{\bf m}_\perp,n},
(A_j)_{\overline{m'_j-1},{\bf m}_\perp,n}],
\label{newvar1}
\end{equation}
and the third term is the $\delta^{(0)}$ variation of a new
interaction
term
\begin{equation}
I^{(2)}_{\text{m}}=-{i\gamma^j\over2\Delta}(ig\Delta)^2
\sum_{m_j'=m_j+1}^{m_j+M}\sum_{m_j''=m'_j+1}^{m'_j+M-1}
(-1)^{m_j+m_j''}
(A_j)_{\overline{m'_j-1},{\bf m}_\perp,n}
(A_j)_{\overline{m''_j-1},{\bf m}_\perp,n}
\psi_{m''_j,\overline{\bf m}_\perp,
\overline{n}^{\vphantom{\prime}}}.
\label{secondspace}
\end{equation}

Similarly, we find the new variation of $A_0$,
\begin{mathletters}
\label{newvar2}
\begin{eqnarray}
n\ge1:\quad\delta^{(1)}(A_0)_{{\bf m},\overline{n-1}}
&=&{ig\over2}[\delta\omega_{{\bf m},n}+\delta\omega_{{\bf m},n-1},
(A_0)_{{\bf m},\overline{n-1}}],\\
n=0:\quad  \delta^{(1)}(A_0)_{{\bf m},\overline{-1}}
&=&{ig\over2}[\delta\omega_{{\bf m},0}+\delta\omega_{{\bf m},-1},
(A_0)_{{\bf m},\overline{-1}}]\nonumber\\&&
\mbox{}+{ig\over4}
[\delta\omega_{{\bf m},0}-\delta\omega_{{\bf m},-1},
(A_0)_{{\bf m},\overline{-1}}],
\end{eqnarray}
\end{mathletters} and a new interaction term involving the
scalar potential,
\begin{equation}
K^{(2)}_{\text{m}}=-2{i\gamma^0\over h}(igh)^2
\sum_{n'=0}^n\!\!{}'\sum_{n''=0}^{n'}\!\!{}''
(-1)^{n+n'}
(A_0)_{{\bf m},\overline{n'-1}}(A_0)_{{\bf m},\overline{n''-1}}
\psi_{\overline{\bf m},n''}.
\label{secondtime}
\end{equation}
Here the double prime on the second summation sign
 means that both the first
and last terms are counted with half weight, and that if the upper
limit is zero, the sum vanishes.

Of course, the $\delta^{(1)}$ variations (\ref{newvar1})  and
(\ref{newvar2}) precisely reduce to the usual non-Abelian
transformations
of a gauge field in the continuum, and the new interaction terms
(\ref{secondspace}) and (\ref{secondtime}) vanish in the continuum
limit.

This iterative procedure continues indefinitely.  Fortunately, it
exponentiates.  Perhaps the easiest way to express $I^{(N)}$, where
$N$ is the order in $g\Delta A$, is by an inductive formula.
Let us adopt a shorthand notation for ease of writing the results.
Let $I=\gamma^j I_j$, and because $I_j$ depends locally on all
coordinates
except $m_j$, let us suppress those other coordinates,
call $m_j=m$, and simply
write $I_m$ to stand for $(I_j)_{m_j,{\bf m}_\perp,n}$.
Then we find
\begin{equation}
N\ge1:\quad I_m^{(N)}={1\over2}\sum_{m'=m+1}^{m+M}
(-1)^{m+m'}\sum_{k=1}^N{1\over k!}(-ig\Delta {\cal A}_{m'})^k
I_{m'}^{(N-k)},
\label{induct}
\end{equation}
where
\begin{equation}
{\cal A}_m=(A_j)_{\overline{m_j-1},{\bf m}_\perp,n}
\label{cala}
\end{equation}
 and we define
$I_m^{(0)}=-2i\psi_{m_j,\overline{\bf m}_\perp,\overline{n}}/\Delta$.
The gauge transformations of the potentials are given by
nested commutators:
\begin{mathletters}
\label{secvar1}
\begin{eqnarray}
k\ne1:\quad \delta^{(k)}{\cal A}_m&=&{(ig\Delta)^k\over\Delta}
{{\cal B}_k\over k!}{\underbrace{[\cdots
[\delta\omega_m-\delta\omega_{m-1},{\cal A}_m],\cdots,{\cal A}_m]}
_{\text{$k$ nested commutators}}},\\
\delta^{(1)}{\cal A}_m&=&{ig\over2}[\delta\omega_m
+\delta\omega_{m-1},{\cal A}_m],
\end{eqnarray}
\end{mathletters}
where ${\cal B}_k$ is the $k$th Bernoulli number.  The required
covariance statement
\begin{equation}
\delta_\psi I^{(N)}_m+\sum_{k=0}^N\delta^{(k)}I^{(N-k+1)}_m
=ig\delta\omega_mI_m^{(N)}
\end{equation}
is proved in the Appendix of \cite{nagt}.  By summing (\ref{induct})
over all $N$ we derive the sum equation satisfied by the full
interaction
term involving the vector potential, $I=\sum_{N=1}^\infty I^{(N)}$:
\begin{equation}
I_m={1\over2}\sum_{m'=m+1}^{m+M}(-1)^{m+m'}
\left(e^{-ig\Delta{\cal A}_{m'}}
-1\right)\left(I_{m'}^{\vphantom{()}}+I_{m'}^{(0)}\right),
\end{equation}
from which a difference equation can be immediately derived:
\begin{equation}
I_m+e^{ig\Delta{\cal A}_m}I_{m-1}={2i\over\Delta}\left(1-
e^{ig\Delta{\cal A}_m}\right)
\psi_{m_j,\overline{\bf m}_\perp,\overline{n}}.
\end{equation}
Of course, $I$ transforms in the required manner
($\delta_A=\sum_{k=0}^\infty \delta^{(k)}$)
\begin{equation}
\delta_\psi I_m+\delta_A I_m-\delta^{(0)}I^{(1)}_m=
ig\delta\omega_m I_m
\end{equation}
where the third variation on the left-hand side is just that
necessary
to cancel the noncovariance of the free Dirac equation (\ref{dirac}).

In just the same way we can derive analogous expressions for the full
interactions and transformations of the scalar potential.
Again we suppress all but the time coordinate $n$, and write
the interaction term as $K\to\gamma^0 K$
in what follows.  The only complication has to do with the initial
conditions.  The order $N$ interaction term is given by
\begin{equation}
K_n^{(N)}=-\sum_{n'=1}^n(-1)^{n+n'}\sum_{k=1}^N{(-igh)^k\over k!}
{\cal A}^k_{n'}K^{(N-k)}_{n'}+{(igh)^N\over 2^N N!}(-1)^n
{\cal A}_0^NK^{(0)}_0,
\label{induct2}
\end{equation}
where $K^{(0)}_n=-2i\psi_{\overline{\bf m},n}/h$.  The required
transformation law
\begin{equation}
\delta_\psi K^{(N)}_n+\sum_{k=0}^N\delta^{(k)}K^{(N-k+1)}_n
=ig\delta\omega_nK_n^{(N)}
\end{equation}
 is satisfied provided the transformations are given by
\begin{mathletters}
\label{secvar2}
\begin{eqnarray}
k\ne1:\quad \delta^{(k)}{\cal A}_n&=&{(igh)^k\over h}
{{\cal B}_k\over k!}{\underbrace{[
\cdots[\delta\omega_n
-\delta\omega_{n-1},{\cal A}_n],\cdots,{\cal A}_n]}
_{\text{$k$ nested commutators}}},\quad n\ne0,\\
\delta^{(1)}{\cal A}_n&=&{ig\over2}[\delta\omega_n
+\delta\omega_{n-1},{\cal A}_n],
\quad n\ne0,\\
k\ne1:\quad \delta^{(k)}{\cal A}_0&=&{(-igh)^k\over h}
{{\cal B}_k\over2^k k!}{\underbrace{[
\cdots[\delta\omega_0-\delta\omega_{-1},{\cal A}_0],\cdots,{\cal
A}_0]}
_{\text{$k$ nested commutators}}},\label{invar1}\\
\delta^{(1)}{\cal A}_0&=&{ig\over2}[\delta\omega_0
+\delta\omega_{-1},{\cal A}_0]
-ig{{\cal B}_1\over2}[\delta\omega_0-\delta\omega_{-1},{\cal A}_0].
\label{invar2}
\end{eqnarray}
\end{mathletters}
$\!\!$A sum and difference equation may be immediately derived for
the full
interaction term involving the scalar potential
$K_n=\sum_{N=1}^\infty
K_n^{(N)}$:
\begin{equation}
K_n=-\sum_{n'=1}^n(-1)^{n+n'}\left(e^{-igh{\cal A}_{n'}}-1\right)
\left(K^{\vphantom{()}}_{n'}+K^{(0)}_{n'}\right)
+(-1)^n\left(e^{igh{\cal A}_0/2}-1\right)
K_0^{(0)},
\end{equation}
and
\begin{equation}
K_n+e^{igh{\cal A}_n}K_{n-1}={2i\over h}\left(1-e^{igh{\cal
A}_n}\right)
\psi_{\overline{\bf m},n}.
\end{equation}

The full lattice Dirac equation is given by (\ref{dirac}),
(\ref{induct}), and (\ref{induct2}):
\begin{equation}
\gamma^0\left[{i\over h}(\psi_{\overline{\bf m},n+1}-
\psi_{\overline{\bf m},n})+K_{{\bf m},n}\right]
+\gamma^j\left[{i\over\Delta}
(\psi_{m_j+1,\overline{\bf m}_\perp,\overline{n}}-
\psi_{m_j,\overline{\bf m}_\perp,\overline{n}})
+(I_j)_{{\bf m},n}\right]
+\mu\psi_{\overline{\bf m},\overline{n}}=0.
\label{fulldirac}
\end{equation}
We have already emphasized that this equation gives
$\psi_{{\bf m},n+1}$ in terms of fields at time $n$ and earlier,
so that this difference equation may be solved by time stepping
through the lattice.

We have given so much detail on the Dirac equation because this
is the hard part of the calculation.  The crucial point is that
we have now determined not merely the interaction terms in the
Dirac equation, but the precise form of the gauge transformations
on the lattice.  We must now use that information to construct the
gauge sector of the theory.  (If one starts directly with the
gauge sector, it is much less clear how to proceed.)

Let us assume that we can construct a gauge-covariant field strength
from the potentials.  (That construction will be given in the
next two sections.)  That is, we have a field strength defined on
the lattice which transforms under an infinitesimal gauge
transformation according to
\begin{equation}
\delta (F_{\mu\nu})_{\overline{\text{m}}}=
ig[\delta\omega_{\text{m}},(F_{\mu\nu})_{\overline{\text{m}}}].
\label{fsxfr}
\end{equation}
(This is precisely analogous to (\ref{local}).)
This is a difference equation for the transformations of the
field strength at the lattice sites.  Because we require boson
fields to be periodic, we must henceforward assume that $M$ is odd.
(Recall the periodicity condition (\ref{periodic}).)  For the
inital condition we adopt the analog of (\ref{initial}),
\begin{equation}
\delta(F_{0i})_{\overline{\bf m},0}
={ig\over2}[\delta\omega_{{\bf m},0}+
\delta\omega_{{\bf m},-1},(F_{0i})_{\overline{\bf m},0}].
\end{equation}

Our starting point for describing gauge fields coupled to a fermionic
current is the ``free'' Yang-Mills equation
\begin{equation}
{1\over h}\sum_{\nu\ne\mu}\left[
(F^{\mu\nu})_{m_\nu+1,\overline{m}_\mu,\overline{\text{m}}_\perp}
-(F^{\mu\nu})_{m_\nu,\overline{m}_\mu,
\overline{\text{m}}_\perp}\right]
=(j^\mu)_{\text{m}}.
\label{freeym}
\end{equation}
Here, for ease of notation, we have not destinguished between the
space and time lattice: $h$ or $\Delta$ should be used as
appropriate.
  A gauge covariant
candidate for the fermionic current is ($T$ is the generator of the
gauge group)
\begin{equation}
j^\mu_{\text{m}}=g\overline\psi_{\overline{\text{m}}}
T\gamma^\mu\psi_{\overline{\text{m}}}^{\vphantom{\dagger}},
\end{equation}
because, then, according to (\ref{local}), the right side of
(\ref{freeym}) transforms covariantly:
\begin{equation}
\delta j^\mu_{\text{m}}=ig[\delta\omega_{\text{m}},j^\mu_{\text{m}}].
\end{equation}
It is the difference in (\ref{freeym}) that spoils gauge covariance.
But now it is immediately obvious that the previous analysis for
the Dirac equation applies.  Multiplication simply gets replaced by
commutation.  We simply must add to the left side of
(\ref{freeym})
the interaction term
\begin{equation}
{\cal I}^\mu=\sum_{\nu\ne\mu}{\cal K}^\nu,\quad {\cal
K}^\nu_{\text{m}}
={\cal K}_n
\end{equation}
(that is, in ${\cal K}^\nu$ the only nonlocal dependence is in
$m_\nu=n$),
where ${\cal K}^\nu$ satisfies the difference equation
\begin{equation}
{\cal K}_ne^{igh{\cal A}_n}+e^{igh{\cal A}_n}{\cal K}_{n-1}
=-{2\over h}\left[e^{igh{\cal A}_n},(F^{\mu\nu})_{m_\nu,
\overline{\text m}_\perp}\right].
\end{equation}
The solution to this difference equation can be
immediately written down by
comparison with the corresponding terms in the Dirac equation
(multiplication being replaced by commutation)
and is given explicitly in \cite{review,nagt}.
In particular, the inductive formul\ae\ are
\begin{equation}
{\cal K}_m^{(N)}={1\over2}\sum_{m'=m+1}^{m+M}
(-1)^{m+m'}\sum_{k=1}^N{(-ig\Delta)^k\over k!}
{\underbrace{[{\cal A}_{m'},[{\cal A}_{m'},[\cdots,[{\cal A}_{m'},
{\cal K}_{m'}^{(N-k)}]\cdots]]]}_{\text{$k$ nested commutators}}},
\end{equation}
for the $\nu=j$ interaction, and
\begin{eqnarray}
{\cal K}_n^{(N)}&=&-\sum_{n'=1}^{n}
(-1)^{n+n'}\sum_{k=1}^N{(-igh)^k\over k!}
{\underbrace{[{\cal A}_{n'},[{\cal A}_{n'},[\cdots,[{\cal A}_{n'},
{\cal K}_{n'}^{(N-k)}]\cdots]]]}_{\text{$k$ nested commutators}}}
\nonumber\\
&&\mbox{}+{(igh)^N\over 2^NN!}(-1)^n
{\underbrace{[{\cal A}_0,[{\cal A}_0,[\cdots,[{\cal A}_0,
{\cal K}_{0}^{(0)}]\cdots]]]}_{\text{$N$ nested commutators}}},
\end{eqnarray}
for the $\nu=0$ interaction.

\section{Iterative Construction of the Field Strength Tensor}
The finite-element prescription for the construction of the
field strength in terms of potentials, in lowest order, is simply
the lattice curl:
\begin{equation}
(F^{(0)}_{\mu\nu})_{\overline{\text{m}}}
={1\over h}[(A_\nu)_{m_\mu+1,\overline{\text{m}}_\perp}-
(A_\nu)_{m_\mu,\overline{\text{m}}_\perp}].
\label{curl}
\end{equation}
(In this section, for simplicity, we set $h=\Delta$; the distinction
can be readily recovered if desired.)
Of course, this is invariant under the transformation $\delta^{(0)}$,
(\ref{zerovar}).  Next, we apply the transformation $\delta^{(1)}$,
(\ref{newvar1}) and (\ref{newvar2}); remarkably enough, we are
able to write the result in the form
\begin{equation}
\delta^{(1)}(F^{(0)}_{\mu\nu})_{\overline{\text{m}}}=
ig[\delta\omega_{\text{m}},
(F^{(0)}_{\mu\nu})_{\overline{\text{m}}}]
-\delta^{(0)}(F^{(1)}_{\mu\nu})_{\overline{\text{m}}},
\label{firstcov}
\end{equation}
that is, the difference from the required covariance
(\ref{fsxfr}) is a
total variation of a new interaction term.  To present this
interaction most simply, let us introduce some further notation.
We use the  averaged potential (\ref{cala}), which we write here as
\begin{equation}
{\cal
A}^\nu_{klmn}={1\over2}\left(A^\nu_{klmn}+A^\nu_{klmn-1}\right),
\end{equation}
where henceforward we order the coordinate indices $\kappa$,
$\lambda$, $\mu$, $\nu$. (Here $n=m_\nu$ is not necessarily the
time coordinate.)
 In terms of this, we define a ``field
strength'' variable local in the transverse coordinates
($\kappa$, $\lambda$)
\begin{equation}
({\cal F}^{(0)}_{\mu\nu})_{klmn}={1\over h}
\left[({\cal A}_\nu)_{klm+1n+1}
-({\cal A}_\nu)_{klmn+1}-({\cal A}_\mu)_{klm+1n+1}
+({\cal A}_\mu)_{klm+1n}\right].
\end{equation}
Under the $\delta^{(0)}$ variation (\ref{zerovar}),
\begin{equation}
\delta^{(0)}({\cal A}_\nu)_{klmn}={1\over h}(\delta\omega_{klmn}
-\delta\omega_{klmn-1}),
\end{equation}
${\cal F}^{(0)}_{\mu\nu}$ is invariant.

In terms of these new variables, the lowest-order field strength is
obtained simply by averaging in the transverse coordinates:
\begin{equation}
(F^{(0)}_{\mu\nu})_{\overline{\text{m}}}=
({\cal F}^{(0)}_{\mu\nu})_{\overline{k}\overline{l}mn}.
\label{fs0}
\end{equation}
Because the first-order variation is given by (\ref{newvar1})
and (\ref{newvar2}) (except for ${\cal A}_0$ at the initial time),
\begin{equation}
\delta^{(1)}({\cal A}_\nu)_{klmn}={ig\over2}[\delta\omega_{klmn}
+\delta\omega_{klmn-1},({\cal A}_\nu)_{klmn}],
\end{equation}
the  interaction term found in (\ref{firstcov}) can now be written as
\begin{eqnarray}
(F^{(1)}_{\mu\nu})_{\overline{\text{m}}}
&=&({\cal F}^{(1)}_{\mu\nu})_{\overline{k}\overline{l}mn}
-{igh\over4}\big\{
[({\cal A}_\kappa)_{k+1lmn},({\cal F}^{(0)}_{\mu\nu})_{k+1lmn}]
+[({\cal A}_\lambda)_{kl+1mn},
({\cal F}^{(0)}_{\mu\nu})_{kl+1mn}]\nonumber\\
&&\mbox{}+{1\over2}
[({\cal A}_\kappa)_{k+1l+1mn},({\cal F}^{(0)}_{\mu\nu})_{k+1l+1mn}]
+{1\over2}[({\cal A}_\lambda)_{k+1l+1mn},({\cal
F}^{(0)}_{\mu\nu})_{k+1l+1mn}]\nonumber\\
&&\mbox{}+{1\over2}[({\cal A}_\kappa)_{k+1lmn},
({\cal F}^{(0)}_{\mu\nu})_{k+1l+1mn}]
+{1\over2}[({\cal A}_\lambda)_{kl+1mn},
({\cal F}^{(0)}_{\mu\nu})_{k+1l+1mn}]
\big\},
\label{fofs}
\end{eqnarray}
where
\begin{eqnarray}
({\cal F}^{(1)}_{\mu\nu})_{klmn}&=&{ig\over2}\big\{
[({\cal A}_\mu)_{klm+1n},({\cal A}_\mu)_{klm+1n+1}]-
[({\cal A}_\nu)_{klmn+1},({\cal A}_\nu)_{klm+1n+1}]\nonumber\\
&&\mbox{}-
[({\cal A}_\mu)_{klm+1n+1},({\cal A}_\nu)_{klm+1n+1}]+
[({\cal A}_\mu)_{klm+1n},({\cal A}_\nu)_{klmn+1}]\nonumber\\
&&\mbox{}-
[({\cal A}_\mu)_{klm+1n},({\cal A}_\nu)_{klm+1n+1}]
-[({\cal A}_\mu)_{klm+1n+1},({\cal A}_\nu)_{klmn+1}]\big\}.
\label{core}
\end{eqnarray}
It will be noted that (\ref{core}) is exactly the form of the
first-order
field strength found in two dimensions \cite{nagt}.

We continue, by finding the second-order field strength from
the covariance equation
\begin{equation}
\delta^{(2)}( F^{(0)}_{\mu\nu})_{\overline{\text{m}}}+
\delta^{(1)}( F^{(1)}_{\mu\nu})_{\overline{\text{m}}}+
\delta^{(0)}( F^{(2)}_{\mu\nu})_{\overline{\text{m}}}=
ig[\delta\omega_{\text{m}},
( F^{(1)}_{\mu\nu})_{\overline{\text{m}}}].
\end{equation}
Here ($n>0$)
\begin{equation}
\delta^{(2)}({\cal A}_\nu)_{klmn}
=-{g^2h\over12}[[\delta\omega_{klmn}-
\delta\omega_{klmn-1},({\cal
A}_\nu)_{klmn}],
({\cal A}_\nu)_{klmn}],
\end{equation}
from (\ref{secvar1}) and (\ref{secvar2}).
The result of a long calculation is remarkably simple:
\begin{eqnarray}
(F^{(2)}_{\mu\nu})_{\overline{\text{m}}}
&=&({\cal F}^{(2)}_{\mu\nu})_{\overline{k}\overline{l}mn}-
{igh\over4}\big\{
[({\cal A}_\kappa)_{k+1lmn},({\cal F}^{(1)}_{\mu\nu})_{k+1lmn}]
+[({\cal A}_\lambda)_{kl+1mn},
({\cal F}^{(1)}_{\mu\nu})_{kl+1mn}]\nonumber\\
&&\quad\mbox{}+{1\over2}
[({\cal A}_\kappa)_{k+1l+1mn},({\cal F}^{(1)}_{\mu\nu})_{k+1l+1mn}]
+{1\over2}[({\cal A}_\lambda)_{k+1l+1mn},({\cal
F}^{(1)}_{\mu\nu})_{k+1l+1mn}]\nonumber\\
&&\quad\mbox{}+{1\over2}[({\cal A}_\kappa)_{k+1lmn},
({\cal F}^{(1)}_{\mu\nu})_{k+1l+1mn}]
+{1\over2}[({\cal A}_\lambda)_{kl+1mn},({\cal
F}^{(1)}_{\mu\nu})_{k+1l+1mn}]
\big\}\nonumber\\
&&-{g^2h^2\over8}\big\{[({\cal A}_\lambda)_{kl+1mn},
[({\cal A}_\lambda)_{kl+1mn},
({\cal F}_{\mu\nu}^{(0)})_{kl+1mn}]]\nonumber\\
&&\quad\mbox{}+
[({\cal A}_\kappa)_{k+1lmn},
[({\cal A}_\kappa)_{k+1lmn},({\cal F}_{\mu\nu}^{(0)})_{k+1lmn}]]
\nonumber\\
&&\quad\mbox{}+{1\over2}[({\cal A}_\kappa)_{k+1l+1mn},
[({\cal A}_\kappa)_{k+1l+1mn},({\cal F}_{\mu\nu}^{(0)})_{k+1l+1mn}]]
\nonumber\\
&&\quad\mbox{}+{1\over2}[({\cal A}_\lambda)_{k+1l+1mn},
[({\cal A}_\lambda)_{k+1l+1mn},({\cal F}_{\mu\nu}^{(0)})_{k+1l+1mn}]]
\nonumber\\
&&\quad\mbox{}+{1\over2}[({\cal A}_\kappa)_{k+1lmn},
[({\cal A}_\kappa)_{k+1lmn},({\cal F}_{\mu\nu}^{(0)})_{k+1l+1mn}]]
\nonumber\\
&&\quad\mbox{}+{1\over2}[({\cal A}_\lambda)_{kl+1mn},
[({\cal A}_\lambda)_{kl+1mn}
,({\cal F}_{\mu\nu}^{(0)})_{k+1l+1mn}]]\nonumber\\
&&\quad\mbox{}+[({\cal A}_\lambda)_{kl+1mn},
[({\cal A}_\kappa)_{k+1l+1mn},({\cal F}_{\mu\nu}^{(0)})_{k+1l+1mn}]]
\nonumber\\
&&\quad\mbox{}+
[({\cal A}_\kappa)_{k+1lmn},
[({\cal A}_\lambda)_{k+1l+1mn},({\cal F}_{\mu\nu}^{(0)})_{k+1l+1mn}]]
\big\},
\label{sofs}
\end{eqnarray}
where
\begin{eqnarray}
({\cal F}^{(2)}_{\mu\nu})_{klmn}&=&
-{g^2h\over12}\big\{[B_1,[B_1,C_1]]-[C_1,[C_1,B_1]]
-[B_1,[B_1,C_0]]+[C_1,[C_1,B_0]]\nonumber\\&&\mbox{}+
2[B_1,[B_0,C_1]]-2[C_1,[C_0,B_1]]+
2[B_0,[B_1,C_1]]-2[C_0,[C_1,B_1]]\nonumber\\&&\mbox{}-
2[B_1,[B_0,C_0]]+2[C_1,[C_0,B_0]]+
4[B_0,[B_1,C_0]]-4[C_0,[C_1,B_0]]\nonumber\\&&\mbox{}
+[B_0,[B_0,C_1]]-[C_0,[C_0,B_1]]-
[B_0,[B_0,C_0]]+[C_0,[C_0,B_0]]\nonumber\\&&\mbox{}
-[B_0,[B_0,B_1]]+[C_0,[C_0,C_1]]-
[B_1,[B_0,B_1]]+[C_1,[C_0,C_1]]\big\}.
\end{eqnarray}
Here we have adopted one final bit of condensed notation:
\begin{equation}
B_1=({\cal A}_\mu)_{klm+1n+1},\quad
B_0=({\cal A}_\mu)_{klm+1n},\quad
C_1=({\cal A}_\nu)_{klm+1n+1},\quad
C_0=({\cal A}_\nu)_{klmn+1}.
\label{abbrev}
\end{equation}

One additional explicit term in ${\cal F}$, for ${\cal
F}_{\mu\nu}^{(3)}$,
was given in \cite{nagt}.

It is quite clear that this process may be continued indefinitely.
We will now see that these terms may be summed exactly.

\section{Exact Field Strength}

At first glance, the pattern for the construction of the
${\cal F}_{\mu\nu}=
\sum_{N=1}^\infty {\cal F}_{\mu\nu}^{(N)}$ seems a bit mysterious.
However, upon a bit of inspection, the terms given here and in
\cite{nagt} are entirely reproduced by the expansion in $h$ of the
following path-ordered product of exponentials (link variables)
 around the finite element (plaquette):
\begin{equation}
e^{-ih^2g({\cal F}_{\mu\nu})_{klmn}}=Pe^{-ig\oint A\cdot dl}
=e^{-ihgB_0}e^{-ihgC_1}e^{ihgB_1}e^{ihgC_0}.
\label{poexp}
\end{equation}
This is precisely the form conventionally used in lattice gauge
theory (implicitly in Wilson's original paper \cite{wilson},
and explicitly, for example, in \cite{rothe}).
Of course, here, the appropriately averaged finite-element potentials
are employed.\footnote{That this is nontrivial can be seen from
the remarks in \cite{rothe}: ``This identification of $A_\mu(n)$
with the vector potential is only strictly correct in the
continuum limit,'' and that only as the lattice spacing
goes to zero do the link variables transform  equivalently to
(\ref{linktrans}).}
  That this is exactly correct emerges immediately when
we recognize the effect of the gauge transformations on the link
operators ($\delta=\sum_{N=0}^\infty\delta^{(N)}$):
\begin{equation}
\delta e^{-igh({\cal A}_{\nu})_{klmn+1}}
=ig\left\{\delta\omega_{klmn} e^{-igh({\cal A}_{\nu})_{klmn+1}}
-e^{-igh({\cal A}_{\nu})_{klmn+1}}\delta\omega_{klmn+1}\right\},
\label{linktrans}
\end{equation}
and the corresponding adjoint equation. The proof is given
in the Appendix.
It then follows immediately from (\ref{poexp}) that
\begin{equation}
\delta({\cal F}_{\mu\nu})_{klmn}
=ig[\delta\omega_{klmn},({\cal F}_{\mu\nu})_{klmn}].
\end{equation}

Now, we immediately recognize the structure of the full
field strength $F_{\mu\nu}=\sum_{N=0}^\infty F_{\mu\nu}^{(N)}$,
the first three terms of which are given in
 (\ref{fs0}), (\ref{fofs}), and (\ref{sofs}):
\begin{eqnarray}
(F_{\mu\nu})_{\overline{k}\overline{l}\overline{m}\overline{n}}
&=&{1\over4}\bigg\{({\cal F}_{\mu\nu})_{klmn}
+ e^{-igh({\cal A}_{\lambda})_{kl+1mn}}
({\cal F}_{\mu\nu})_{kl+1mn}
 e^{igh({\cal A}_{\lambda})_{kl+1mn}}\nonumber\\&&\mbox{}+
 e^{-igh({\cal A}_{\kappa})_{k+1lmn}}
({\cal F}_{\mu\nu})_{k+1lmn}
 e^{igh({\cal A}_{\kappa})_{k+1lmn}}\nonumber\\
&&\mbox{}+{1\over2} e^{-igh({\cal A}_{\lambda})_{kl+1mn}}
 e^{-igh({\cal A}_{\kappa})_{k+1l+1mn}}
({\cal F}_{\mu\nu})_{k+1l+1mn}
 e^{igh({\cal A}_{\kappa})_{k+1l+1mn}}
 e^{igh({\cal A}_{\lambda})_{kl+1mn}}\nonumber\\&&\mbox{}+
{1\over2}e^{-igh({\cal A}_{\kappa})_{k+1lmn}}
 e^{-igh({\cal A}_{\lambda})_{k+1l+1mn}}
({\cal F}_{\mu\nu})_{k+1l+1mn}
 e^{igh({\cal A}_{\lambda})_{k+1l+1mn}}
 e^{igh({\cal A}_{\kappa})_{k+1lmn}}\bigg\}.\nonumber\\
\end{eqnarray}
That $F_{\mu\nu}$ transforms covariantly according to (\ref{fsxfr})
follows immediately from (\ref{linktrans}).
That is, the field strength is the average of the
``transversely-local'' field strength ${\cal F}_{\mu\nu}$ over the
four corners of the transverse finite element, with gauge factors
inserted appropriately to gauge-transform the field strength back
to the origin $(klmn)$.

Finally, we note that at the initial time $F_{\mu\nu}$ involves
$({\cal A}_0)_{{\bf m},1}$ and not
$({\cal A}_0)_{{\bf m},0}$, so that the initial variations
(\ref{invar1}) and (\ref{invar2}) never come into play.

\section{Conclusions}

We have thus finished the task of giving a complete set of
gauge-invariant
equations of motion for Yang-Mills fields coupled to fermions.
This should enable us to apply finite-element methods of solution
which have so far been applied to scalar field theories \cite{bm}
and to electrodynamics \cite{mmsb,mmsb2,twodqed}.  New methods
are being developed as well \cite{hamm}.  It is therefore
reasonable to  hope
that we may be able to use this  technique to shed new light
on difficult problems in non-Abelian theories such as QCD.

\section*{Acknowledgments}
I thank the UK PPARC for the award of a Senior Visiting
Fellowship and Imperial College for its hospitality.
I further  thank the US Department of Energy and the
University of Oklahoma College of Arts and Sciences
 for partial financial support of this
research. I am grateful to Tai Wu for useful conversations
and George Kalbfleisch for support and encouragement.

\appendix
\section*{Proof of gauge transformations of link operators}
The proof of (\ref{linktrans}) follows essentially from the proof
of covariance given in the Appendix of\cite{nagt}.
We have, for example, from (\ref{secvar1}) (we are using the notation
(\ref{abbrev}) and only display the values of the $\mu$, $\nu$
indices) ($n>0$)
\begin{equation}
\delta^{(k)}e^{ighB_0}
=\sum_{N=0}^\infty{(igh)^N\over N!}\sum_{l=0}^{N-1}
B_0^l{(igh)^k\over h}{{\cal B}_k\over k!}{\underbrace{
[\cdots[\delta\omega_{m+1n}-\delta\omega_{mn},B_0],\cdots, B_0]}
_{\text{$k$ nested commutators}}}
B_0^{N-l-1}, \quad k\ne1,
\end{equation}
and
\begin{equation}
\delta^{(1)}e^{ighB_0}=\sum_{N=0}^\infty{(igh)^N\over N!}
\sum_{l=0}^{N-1}B_0^l{ig\over2}[\delta\omega_{m+1n}
+\delta\omega_{mn},
B_0]B_0^{N-l-1},
\end{equation}
so the total variation of the link variable is
\begin{eqnarray}
\delta e^{ighB_0}&=&\sum_{k=0}^\infty\delta^{(k)}e^{ighB_0}
=ig[\delta\omega_{mn},e^{ighB_0}]\nonumber\\&&\mbox{}\!\!\!
+{1\over h}\sum_{k=0}^\infty(-1)^k
\sum_{N=1}^\infty\sum_{l=0}^{N-1}
B_0^l{(igh)^{N+k}\over N!}{{\cal B}_k\over k!}{\underbrace{
[\cdots[\delta\omega_{m+1n}-\delta\omega_{mn},B_0],\cdots, B_0]}
_{\text{$k$ nested commutators}}}B_0^{N-l-1}
\label{var}
\end{eqnarray}
The nested commutator is simply given by
\begin{equation}
{\underbrace{{[\cdots[\delta\omega,B],\cdots, B]}}
_{\text{$k$ nested commutators}}}
=\sum_{s=0}^k(-1)^s\left(
\begin{array}{c}k\\s\end{array}\right)B^s\delta\omega B^{k-s}.
\end{equation}
Then the sum on the right hand side of (\ref{var}) reduces to
\begin{eqnarray}
&&{1\over h}\sum_{r=0}^\infty\sum_{p=0}^r(igh)^{r+1} B_0^p
(\delta\omega_{m+1n}-\delta\omega_{mn})B_0^{r-p}\nonumber\\
&&\quad\times\sum_{k=0}^r(-1)^k{{\cal B}_k\over k!}{1\over(r+1-k)!}
\sum_{s=p-r+k}^p(-1)^s\left(\begin{array}{c}k\\s\end{array}
\right)=ig(\delta\omega_{m+1n}-\delta\omega_{mn})e^{ighB_0},
\end{eqnarray}
where the last two summations over $s$ and $k$ are carried out
using identity (A3) of \cite{nagt}.
Thus we have established the desired result:
\begin{equation}
\delta e^{ighB_0}=ig(\delta\omega_{m+1n}e^{igh B_0}
-e^{ighB_0}\delta\omega_{mn}),
\end{equation}
which is the adjoint of (\ref{linktrans}).

\end{document}